\documentclass[a4paper,11pt]{article}
\usepackage{pos}
\usepackage{amsmath}
\numberwithin{equation}{section}
\usepackage{mathtools}
\usepackage{mathpazo}
\usepackage{multirow}
\usepackage{graphicx}
\usepackage{ifpdf}
\usepackage{braket}
\usepackage{cancel}

\usepackage[normalem]{ulem}

\usepackage{accents}

\title{Generalised Cartan Geometry}

\author*[a]{David Osten}

\affiliation[a]{Institute for Theoretical Physics (IFT), \\
University of Wroc\l aw \\
pl. Maxa Borna 9, 50-204 Wroc\l aw, Poland}

\emailAdd{david.osten@uwr.edu.pl}

\abstract{This talk introduces a Cartan-geometric framework for generalised geometries governed by a differential graded Lie algebra. In contrast to ordinary Cartan geometry, the tangent bundle is extended and qu both a global duality group and a local gauge group. This framework provides a systematic construction of generalised connections and their torsion and curvature tensors for generic generalised geometries. We also review the realisation of these algebraic structures on the phase space of branes in M-theory. 
}

\FullConference{Proceedings of the Corfu Summer Institute 2025 "School and Workshops on Elementary Particle Physics and Gravity" (CORFU2025)\\
27 April - 28 September, 2025\\
Corfu, Greece\\}

\begin{document}
\maketitle

\section{Introduction}\label{chap:Intro}

Geometry is key in modern physics, from the spacetime geometry of general relativity to the gauge structures underlying particle and string theory. A central lesson from string and M-theory is that the relevant geometry is often larger than ordinary differential geometry. For the bosonic string, the phase space exhibits the $\mathrm{O}(d,d)$ duality symmetry of T-duality \cite{Duff:1989tf,Tseytlin:1990va,Tseytlin:1990nb,Siegel:1993bj,Siegel:1993th,Siegel:1993xq,Alekseev:2004np}; more generally, higher-dimensional brane $\sigma$-models are naturally organised by exceptional duality groups $E_{d(d)}$ \cite{Duff:1990hn,Sakatani:2016sko,Sakatani:2017vbd,Arvanitakis:2018hfn,Blair:2019tww,Sakatani:2020umt,Berman:2010is,Hatsuda:2012uk,Hatsuda:2012vm,Hatsuda:2013dya,Hatsuda:2020buq,Duff:2015jka,Sakatani:2020iad,Sakatani:2020wah,Strickland-Constable:2021afa,Osten:2021fil,Arvanitakis:2021wkt,Hatsuda:2022zpi,Osten:2023iwc,Osten:2024mjt}. Making these symmetries manifest leads to generalised geometry and its exceptional extensions, where the tangent bundle is replaced by an extended bundle such as $TM\oplus T^*M$ or its higher-form analogues. This framework unifies diffeomorphisms with gauge transformations of form fields and has become a powerful language for manifestly duality-symmetric theories \cite{Hull:2004in,Hull:2006va,Hull:2007zu,Pacheco:2008ps,Grana:2008yw,Hull:2009mi,Zwiebach:2011rg,Geissbuhler:2013uka,Aldazabal:2013sca,Berman:2013eva,Plauschinn:2018wbo,Berman:2010is,Berman:2011jh,Coimbra:2011ky,Berman:2012vc,Berman:2012uy,Coimbra:2012af,Hohm:2013pua,Lee:2014mla,Hohm:2014qga}, see \cite{Berman:2020tqn,Sterckx:2024vju,Samtleben:2025fta} for recent reviews. Nevertheless, the differential geometry of generalised geometry (in particular, its Riemannian geometry) turns out to be more subtle. In particular, metric compatibility and vanishing torsion do not uniquely determine a generalised Levi-Civita connection, and defining tensorial notions of torsion and curvature is more delicate than in the ordinary case. Existing constructions of generalised curvature often rely on special structures or leave undetermined components that drop out only in certain physical observables \cite{Siegel:1993th,Gualtieri:2007bq,Hohm:2011si,Coimbra:2011nw,Coimbra:2011ky,Hohm:2012mf,Cederwall:2013naa,Garcia-Fernandez:2016ofz,Jurco:2016emw,Cavalcanti:2024uky,Cortes:2025lns}.

Cartan geometry \cite{Cartan1922,Sharpe:1997,cap2009parabolic,Mackenzie_2005,Blaom:2006,crampin2016cartan,Attard:2019pvw,Alexandre2024} provides a natural starting point to approach this. A Cartan connection combines the frame and the $H$-connection into a single object, valued in a 'model' Lie algebra $\mathfrak{g}$ of a Lie group $G$. Its curvature then packages both torsion and Riemann curvature. In ordinary Cartan geometry, this happens as the manifold is viewed locally as a homogeneous space $G/H$ with the key example: $G$ -- Poincare group and $H$ -- Lorentz group.

Our proposal in \cite{Butter:2022iza,Hassler:2024hgq,Hassler:2025rag} is therefore to develop a minimal Cartan-geometric framework adapted to extended geometries. We formulate it on a principal $H$-bundle, replace the usual Lie-algebroid structure by a more general algebroid controlled by a differential graded Lie algebra, and derive the corresponding curvatures and their hierarchy. This provides a unified geometric description of local gauge symmetry, duality covariance, and phase-space structures relevant to double and exceptional field theory. In addition to the analogues of the frame and spin connection, this generalised Cartan connection contains additional fields required for covariance. Its curvature correspondingly decomposes into generalised torsion, the curvature of the generalised spin connection, and higher curvatures. These objects are naturally organised into a tensor hierarchy, reflecting the structure already familiar from extended field theories.

This framework is motivated both by the phase-space description of string $\sigma$-models and by earlier constructions of covariant generalised curvatures, in particular the work of Pol\'a\v{c}ek and Siegel \cite{Polacek:2013nla}. From our perspective, those constructions are naturally understood as instances of a Cartan-geometric extension of generalised geometry. The advantage of this viewpoint is that it yields a systematic and covariant formulation of connections and curvatures, while also making their role in current algebras and Hamiltonian dynamics transparent.

The paper is organised as follows. In section~\ref{chap:ExtExFT}, we define the class of generalised geometries based on $H\times \mathcal G$. In section~\ref{chap:CurrentAlgebra}, we review the exceptional structure of the brane phase spaces and in section~\ref{chap:Curvature} the main geometric results -- the hierarchy of linearised curvatures of the full generalised Cartan connection, including the higher gauge fields, and their organisation into a tensor hierarchy -- is presented.

\section{Cartan geometry} \label{chap:CartanGeometryOrdinary}

Before turning to its generalisation, let us briefly recall the ordinary Cartan-geometric framework in the form relevant for this article. Our emphasis will be less on global subtleties and more on the conceptual structures that later admit a natural lift to generalised geometry.

\subsection{Cartan connection and curvature}

Cartan geometry provides a way to describe a curved manifold by comparing it, point by point, with a homogeneous model space $G/H$ where $G$ is a Lie group and $H$ a subgroup. The quotient directions describe infinitesimal translations, while $H$ acts as the local isotropy group. In physical language, $H$ plays the role of a local gauge symmetry. Consequently, instead of viewing the tangent space $T_p M$ merely as a vector space, one thinks of it as locally modelled as the quotient algebra $\mathfrak{m} = \mathfrak{g}/ \mathfrak{h}$ where $\mathfrak{g}$ and $\mathfrak{h}$ are the Lie algebra of $G$ and $H$ respectively. 

The standard examples are familiar. If $H=\mathrm{O}(d)$, the model space is Euclidean space and one recovers the geometry underlying metric-compatible affine connections. Similarly, Minkowski space is the homogeneous space $\mathrm{ISO}(1,d-1)/\mathrm{SO}(1,d-1)$, so Lorentzian geometry can be phrased in exactly the same language. Cartan geometry therefore packages frame field and gauge connection into a single object adapted to the symmetry structure of the problem.

Since our later applications are local, we work on a coordinate patch $U\subset M$ and suppress most global issues; for a more complete treatment, see \cite{cap2009parabolic}. The basic ingredients are as follows.

First, one chooses a Lie group $H$ with Lie algebra $\mathfrak h$, together with a principal $H$-bundle
\begin{equation}
    \pi:P\to M,
    \qquad
    \text{locally } P|_U \cong H\times U .
    \label{eq:PrincipalBundleDefinition}
\end{equation}
The central object is the Cartan connection. It is a fibrewise isomorphism
\begin{equation}
    \theta(p):T_pP\to\mathfrak g
    \label{eq:CartanConnectionDefinition}
\end{equation}
which identifies the tangent directions on the bundle with elements of the model algebra $\mathfrak{g}$. The point of this definition is that, after quotienting out the vertical directions generated by $\mathfrak h$, one recovers the identification of $T_xM$ with the translational part $\mathfrak g/\mathfrak h$.

For this interpretation to work, $\theta$ must satisfy two properties. First, it reproduces the fibre direction generators: if $X_\xi$ denotes the fundamental vector field associated with $\xi\in\mathfrak h$, then
\begin{equation}
    \theta(X_\xi)=\xi .
    \label{eq:NormalCartanIdentity}
\end{equation}
Second, it transforms equivariantly under the right action of the gauge group,
\begin{equation}
    R_h^*\theta=\mathrm{Ad}_{h^{-1}}\theta .
    \label{eq:Equivariance}
\end{equation}
This is the geometric origin of the local $H$-gauge symmetry: the decomposition into ``translation'' and ``rotation'' depends on the point in the fibre, and changing that point acts by the adjoint representation of $H$.

To connect with the later phase-space discussion, it is useful to write $\theta$ in a local basis adapted to the splitting into vertical and horizontal directions. We therefore use indices $\mathcal A=(\alpha,a)$ on $TP$ and $\mathcal M=(\mu,m)$ on $\mathfrak g=\mathfrak h\oplus \mathfrak g/\mathfrak h$, and choose a basis of the form $(\nabla_\alpha,\partial_a={e_a}^m\partial_m)$.

Then a general local expression for the Cartan connection is
\begin{equation}
{\theta_{\mathcal A}}^{\mathcal M}(x)=
\begin{pmatrix}
\delta_\alpha{}^\mu & 0 \\
{\omega_a}^{\mu}(x) & {e_a}^m(x)
\end{pmatrix}.
\label{eq:CartanConnectionChoice}
\end{equation}
Its content is exactly what one expects: the non-trivial components are a coframe $e_a{}^m$ and an $\mathfrak h$-valued connection $\omega_a$. In other words, Cartan geometry unifies the vielbein and the gauge connection into a single object. In the present conventions, we keep also the vertical components of $\theta$, which is convenient for the algebraic reformulation below. The \textit{equivariance} property \eqref{eq:Equivariance} means that $e$ and $\omega$ transform as an $\mathfrak{h}$-tensor, i.e.
\begin{equation}
 \nabla_\alpha T_{b ...}^{\gamma ...} \equiv - {f_{\alpha b}}^c T_{c ....}^{\gamma ...} + {f_{\alpha \beta}}^\gamma T_{b ...}^{\beta ...}\,. \label{eq:OrdHTensors} 
\end{equation}
The associated curvature to $\theta$ is the $\mathfrak g$-valued two-form
\begin{equation}
    \Theta=-\mathrm d\theta+\frac12[\theta,\theta],
    \label{eq:CartanCurvatureNormal}
\end{equation}
characteristically combining differential and Lie algebric properties of $\theta$.

Once pulled back to the base manifold, its horizontal components split into the usual torsion and curvature,
\begin{align}
    T &= -\mathrm d e + [\omega,e] \in \mathfrak g/\mathfrak h ,
    \label{eq:TorsionStandard} \\
    R &= -\mathrm d\omega + \frac12[\omega,\omega] \in \mathfrak h .
    \label{eq:CurvatureStandard}
\end{align}
Thus, Cartan curvature combines in one object what in ordinary differential geometry appears as two separate quantities. This is one of the main conceptual advantages of the formalism: torsion and Riemann curvature arise on an equal footing from the curvature of a single connection. In our slightly extended local basis, the remaining components of $\Theta$ encode the Lie algebra structure of $\mathfrak h$ and its action on the tangent directions.

\paragraph{A useful local viewpoint.} For later comparison with generalised geometry, it is helpful to view the above structure from a slightly different angle. One may start from a fibre-independent object $\overline\theta$ of the same block form as in \eqref{eq:CartanConnectionChoice} and then restore the correct $H$-behaviour by twisting it with data intrinsic to the group $H$: the adjoint action on $\mathfrak g$ and the Maurer--Cartan form on the fibre. In this way, the defining properties \eqref{eq:NormalCartanIdentity} and \eqref{eq:Equivariance} are not imposed by hand, but arise naturally from the geometry of the principal bundle itself.

Conceptually, this is useful because it cleanly separates two types of information. The fibre data encode the local symmetry group $H$ and how it acts, while the actual geometry of the base manifold is contained in the frame $e$ and the connection $\omega$. This separation will later reappear in the generalised setting, where it becomes even more important.

\subsection{Realisation of Cartan geometry on the point particle phase space}\label{chap:PointParticle}

Cartan geometry is not only a geometric construction on bundles; it can also be realised directly on the phase space of a point particle. This observation will serve as a prototype for the worldsheet and brane constructions considered later.

Let $x^m$ and $p_m$ denote the canonical coordinates and momenta of a particle moving on $M$, with $\{p_m,x^n\}=-\delta_m^n $. Suppose, in addition, that the phase space carries generators $s_\mu$ of a local $H$-symmetry. Their Poisson brackets take the form
\begin{align}
    \{ s_\mu , s_\nu \} &= {f^\lambda}_{\mu \nu} s_\lambda ,
    \label{eq:CartanGeometryPPGauge}\\
    \{ s_\mu , p_m \} &= {f_{\mu m }}^n p_n ,
    \label{eq:CartanGeometryPPAction}\\
    \{ p_m , p_n \} &= 0 .
    \label{eq:CartanGeometryPPMomenta}
\end{align}
The first relation simply says that the $s_\mu$ generate the Lie algebra $\mathfrak h$. The second expresses that the momenta transform under $H$. Together, these brackets realise the semidirect product structure of the model algebra. In particular, the combined generators $J_{\mathcal M}=(s_\mu,p_m)$ span the analogue of the Euclidean or Poincar\'e algebra on phase space.

The crucial step is now to pass from this ``flat'' algebraic basis to one adapted to the curved geometry. This is achieved by the Cartan connection $J_{\mathcal A}={\theta_{\mathcal A}}^{\mathcal M}J_{\mathcal M}$ or, explicitly,
\begin{equation}
    j_\alpha=s_\alpha,
    \qquad
    j_a={e_a}^m p_m+{\omega_a}^\mu s_\mu .
    \label{eq:CartanGeometryPPNewVariables}
\end{equation}
This is the phase-space counterpart of introducing a vielbein and spin connection. The variables $j_a$ are no longer bare momenta; they are covariantised momenta. They already know about the local frame and the gauge symmetry.

The resulting Poisson algebra is then no longer trivial. The brackets involving $j_\alpha$ continue to encode the action of $H$, while the bracket of two covariant momenta becomes
\begin{equation}
    \{ j_a , j_b \}
    = {\Theta^{\mathcal C}}_{ab} J_{\mathcal C}
    = {T^c}_{ab} j_c + {R^\gamma}_{ab} j_\gamma .
    \label{eq:CartanGeometryOrdCURRENTALGEBRA}
\end{equation}
This is the key conceptual point: the Cartan curvature appears directly as the obstruction to the covariant momenta commuting. In other words, curvature and torsion are encoded in the current algebra of phase-space variables.

\paragraph{Dynamics of the particle.}

This reformulation becomes especially natural once one considers the particle Hamiltonian. For a metric $g$ on $M$, the standard expression is
\begin{equation}
  \mathcal H=\frac12 g^{mn}(x)p_mp_n.
\end{equation}
If one chooses a frame such that $\eta_{ab}=e_a{}^m e_b{}^n g_{mn}$, then in the Cartan basis the Hamiltonian becomes $\mathcal H =\frac12\eta^{ab}\big(j_a-\omega_a^\alpha s_\alpha\big)\big(j_b-\omega_b^\beta s_\beta\big)$. Upon imposing the $H$-generators as first-class constraints, $s_\mu\approx0$, this reduces to
\begin{equation}
  \mathcal H \approx \frac12\eta^{ab}j_aj_b .
\end{equation}
So in the physical subspace the Hamiltonian takes the same form as in flat space, but now written in terms of covariant momenta. The geometry has been absorbed into the current algebra.

The corresponding velocity is $ v^a = {e_m}^a \dot x^m \approx \eta^{ab}j_b$, and its equation of motion becomes
\begin{equation}
    \dot v^a + {\omega_{b,c}}^a v^b v^c = 0 ,
    \label{eq:GeodesicEq}
\end{equation}
which is the geodesic equation in frame variables. Specifically, one obtains the autoparallel equation for $H$-connection $\nabla_a = \partial_a + \omega_a^\alpha \nabla_\alpha$
\begin{equation}
    v^a\nabla_a v^b=0 .
\end{equation}
This gives a clear physical interpretation of the Cartan connection: it is the structure that turns canonical particle motion into covariant motion on a curved background.

\paragraph{Jacobi identities and Bianchi identities.}

The Poisson algebra does more than encode curvature; its Jacobi identities reproduce the consistency conditions of differential geometry. The Jacobi identities that involve the symmetry generators express the closure of the $H$-action. Those involving one $j_\alpha$ and two $j_a$ imply that torsion and curvature transform covariantly under $H$. Finally, the Jacobi identity of three $j_a$ yields the differential Bianchi identities. In this sense, the Bianchi identities are nothing but associativity conditions of the current algebra. This observation is important for the later generalisation: once geometry is encoded in a bracket algebra, many familiar differential identities have algebraic incarnations.

\paragraph{Cartan curvature as a bracket.}

The phase-space picture suggests an alternative but very efficient way to phrase Cartan geometry. Consider $\mathfrak g$-valued fields on $P$ and define the bracket
\begin{equation}
    [V,W]_{L,\mathfrak g}^{\mathcal M}
    =
    V^{\mathcal N} D_{\mathcal N} W^{\mathcal M}
    -
    W^{\mathcal N} D_{\mathcal N} V^{\mathcal M}
    -
    {f^{\mathcal M}}_{\mathcal K\mathcal L}V^{\mathcal K}W^{\mathcal L},
    \label{eq:CartanGeometryLieBracket}
\end{equation}
with $D_{\mathcal M}=(\nabla_\mu,\partial_m)$. In the point-particle realisation, this bracket is simply the Poisson bracket of the corresponding phase-space functions. The Cartan curvature then takes the compact form
\begin{equation}
    \Theta_{\mathcal A\mathcal B}
    =
    -[\theta_{\mathcal A},\theta_{\mathcal B}]_{L,\mathfrak g}
    \in\mathfrak g .
    \label{eq:CartanCurvatureOrdLieBracketRealisation}
\end{equation}
This formula makes the relation between connection, current algebra, and curvature completely explicit. It is also the form that admits the cleanest generalisation later on, once the ordinary Lie bracket is replaced by its generalised counterpart.

To summarise, the Poisson algebra of the covariant phase-space generators reproduces the full local content of Cartan geometry. The components of the extended curvature have a simple interpretation:
\begin{itemize}
    \item $\Theta^\gamma{}_{\alpha\beta}$ encodes the Lie algebra structure of the gauge symmetry $\mathfrak h$;
    \item $\Theta^c{}_{\alpha b}$ describes the action of $\mathfrak h$ on the tangent directions;
    \item $\Theta^{\mathcal M}{}_{ab}$ contains the physical torsion and curvature on $M$.
\end{itemize}
The generalised construction developed later will enlarge this pattern: additional components of the generalised Cartan curvature will appear, and their interpretation will be one of the main outcomes of the paper.

\section{\texorpdfstring{$H \times \mathcal{G}$}{H x G} Generalised Geometry} \label{chap:ExtExFT}

We now introduce the class of extended geometries that will underlie the later Cartan-geometric construction. The basic idea is simple: ordinary generalised geometry makes a duality group $\mathcal G$ manifest, whereas our aim is to incorporate at the same time an additional local symmetry group $H$. In other words, we seek a geometric framework that treats duality covariance and gauge covariance on the same footing.

From the perspective of this paper, this extension should be minimal. We do not enlarge the duality group itself beyond what is needed, but rather build the smallest geometric structure that simultaneously carries the action of $\mathcal G$ and $H$. The resulting bundle will play the role of an extended generalised tangent bundle, and it will come equipped with the analogue of the usual operations of generalised geometry: the tensor hierarchy, a differential, and the generalised Lie derivative.

\subsection{\texorpdfstring{$\mathcal{G}$}{G}-generalised geometry}
We begin with a generalised geometry associated with a duality group $\mathcal G$, such as $\mathrm{O}(d,d)$ or $E_{d(d)}$. Its basic algebraic structure is the tensor hierarchy, a graded vector space
\begin{equation}
    \mathcal T = \mathcal R_1 \oplus \mathcal{R}_2 \oplus ... ,
\end{equation}
where $\mathcal R_p$ are representations of $\mathcal G$ and are treated as vector spaces of degree $p$. Finally, these representations are associated to bundles over the relevant spacetime geometry. The $\mathcal{R}_1$ representation takes a special role here, as the associated bundle is treated as the 'extended' tangent bundle. For example, for $\mathcal{G}=$O$(d,d)$ the extended tangent bundle (or the $\mathcal{R}_1$-bundle) is $TM \oplus T^\star M$ associated to the fundamental $\textbf{2d}$-representation of O$(d,d)$. Also, in order to obtain we formerly extend our $d$-dimensional coordinates $x^\mu$ of $M$ to coordinates $X^{M_1}$.\footnote{We use the index notation $M_p$ for objects in $\mathcal{R}_p$.} In the following, we will introduce a $\mathcal{G}$-covariant constraint that reduces the allowed dependences on these extended coordinates $X^{M_1}$ back to $x^\mu$.

These representations are equipped with two operations: a graded skewsymmetric product $\bullet:\mathcal R_p\otimes \mathcal R_q \longrightarrow \mathcal R_{p+q}$ and a differential $\partial:\mathcal R_p \longrightarrow \mathcal R_{p-1}$.
In components, we write
\begin{align}
    (V\bullet W)^{L_{p+q}} &= {\eta^{L_{p+q}}}_{M_pN_q}\,V^{M_p}W^{N_q},
    \label{eq:EtaDef_rewrite}\\
    (\partial V)^{L_{p-1}} &= {D_{K_p}}^{L_{p-1}M_1}\,\partial_{M_1}V^{K_p},
\end{align}
where the $\eta$- and $D$-symbols encode the algebraic data of the hierarchy. With this data, the tensor hierarchy becomes a differential graded Lie algebra \cite{Bossard:2017aae,Bossard:2018utw,Bossard:2021jix,Cederwall:2021ymp,Cederwall:2025muh,Bossard:2019ksx,Bossard:2021ebg}. The consistency of these operations is controlled by the axioms of the differential graded Lie algebra. For the present discussion, another important one is the section condition
\begin{equation}
    {D_{K_2}}^{L_1M_1}\,\partial_{L_1} \otimes \partial_{M_1}=0,
    \label{eq:SectionCondition_rewrite}
\end{equation}
which restricts the allowed coordinate dependence. Once solved, one recovers an underlying manifold $M$, and the representation $\mathcal R_1$ becomes the generalised tangent bundle $\mathcal R_1 \sim TM \oplus \dots$. On this bundle, one defines the generalised Lie derivative
\begin{equation}
    \mathcal L_V W^{N_1}
    =
    V^{M_1}\partial_{M_1}W^{N_1}
    -
    W^{M_1}\partial_{M_1}V^{N_1}
    +
    Y^{M_1N_1}{}_{K_1L_1}\,\partial_{M_1}V^{K_1}W^{L_1},
    \label{eq:GeneralisedLieDerivative_rewrite}
\end{equation}
with $Y^{M_1N_1}{}_{K_1L_1} = D_{P_2}{}^{M_1N_1}\,\eta^{P_2}{}_{K_1L_1}$. This action of a 'generalised vector' on another generalised vector can be extended to arbitrary 'generalised tensors'.

The algebra of generalised diffeomorphisms, given by the antisymmetrisation of \eqref{eq:GeneralisedLieDerivative_rewrite}, 
\begin{equation}
    [V_1,V_2]_C = \frac12\big(\mathcal L_{V_1}V_2-\mathcal L_{V_2}V_1\big),
\end{equation}
closes on section \cite{Berman:2012vc}. This standard $\mathcal G$-generalised geometry is the starting point of the construction. What we now add is an extra local symmetry group $H$.

\subsection{An extended generalised geometry}\label{sec:extGG}

Let $H$ be a Lie group with Lie algebra $\mathfrak h$. We assume that $\mathfrak h$ acts on the representations $\mathcal R_p$ of the tensor hierarchy. In components, this action is described by matrices $f_{\alpha M_p}{}^{N_p}$ satisfying the representation property
\begin{equation}
    f_{\alpha\beta}{}^\gamma\,f_{\gamma M_p}{}^{N_p}
    =
    -2\,f_{[\alpha|M_p}{}^{K_p}f_{|\beta]K_p}{}^{N_p},
    \label{eq:StructureConstantsRepresentation_rewrite}
\end{equation}
where $f_{\alpha\beta}{}^\gamma$ are the structure constants of $\mathfrak h$. Moreover, we require that this $H$-action preserves the tensor hierarchy data. In other words, the $\eta$- and $D$-symbols of the original $\mathcal G$-geometry must be invariant under $\mathfrak h$. Conceptually, this is what allows the two symmetries to coexist inside a single consistent structure. 

Let us note the central difference to the approach in \cite{Hassler:2023axp,Hassler:2025qhh}. There, not only a duality group $\mathcal{G} = E_{d(d)}$ and a gauge group $H$ was geometrised, but a much larger duality group: $E_{d+n(d+n)}$ for $n=$dim$H$. The geometry, that is presented here shares many features with this approach, in particular will the extended representations presented below be contained in the extended representation of $E_{d+n(d+n)}$. 

\paragraph{Extended representations.}

The first step is to enlarge the ordinary representations $\mathcal R_p$ by adjoining form degrees in $\mathfrak h^*$. The extended analogue of the generalised tangent bundle is
\[ \mathfrak R_1 = \mathfrak h \oplus \mathcal R_1 \oplus (\mathcal R_2\otimes\mathfrak h^*) \oplus (\mathcal R_3\otimes\wedge^2\mathfrak h^*) \oplus \dots . \]
Likewise, for $p>1$ we define
\begin{equation} 
\mathfrak R_p = \oplus_{q\geq 0} \big(\mathcal R_{p+q}\otimes \wedge^q\mathfrak h^*\big). \label{eq:ExtendedRp_rewrite}
\end{equation}

These representations are the minimal extension of the usual tensor hierarchy that carries both the $\mathcal G$-representations and the additional $H$-covariance. Although they are written as infinite sums, they truncate once the form degree in $\mathfrak h^*$ exceeds $\dim\mathfrak h$.

For later purposes, it is also convenient to introduce a degree-zero representation
\begin{equation}
    \mathfrak{R}_0 = ... \oplus (\text{ad}(\mathfrak{h}) \oplus \mathcal{R}_0) \oplus_{q \geq 1} \Big( \mathcal{R}_{q} \otimes \bigwedge\nolimits^{q} \mathfrak{h}^* \Big)\,.
\end{equation}. 
Its main role is to provide a natural home for the generalised spin connection and its higher analogues. We will not need its full structure here, but only the fact that it contains a distinguished subspaces in which the frame- and connection-type fields live.

\paragraph{Extended hierarchy tensors.}

The original $\eta$- and $D$-symbols extend to the enlarged representations $\mathfrak R_p$. Their full component expressions are lengthy \cite{}, but the conceptual point is simple: the ordinary tensor hierarchy operations are promoted so that they also keep track of the $H$-form degree. In particular, they define an extended $Y$-tensor,
\begin{equation}
    Y^{\mathcal K_1\mathcal L_1}{}_{\mathcal M_1\mathcal N_1}
    =
    D_{\mathcal P_2}{}^{\mathcal K_1\mathcal L_1}
    \eta^{\mathcal P_2}{}_{\mathcal M_1\mathcal N_1},
    \label{eq:ExtendedYtensor_rewrite}
\end{equation}
which is the basic building block of the extended generalised Lie derivative.

\paragraph{Section condition and coordinates.}

We associate formal coordinates $X^{\mathcal M_1}$ to the representation $\mathfrak R_1$. The section condition takes the same structural form as before,
\begin{equation}
    D_{\mathcal L_2}{}^{\mathcal M_1\mathcal N_1}\,
    \partial_{\mathcal M_1}\otimes \partial_{\mathcal N_1}
    =0.
\end{equation}
Its natural solution keeps only the coordinates associated with $\mathfrak h$ and the ordinary $\mathcal R_1$-coordinates:
\begin{equation}
    \partial_{\mathcal M_1}
    =
    (\partial_\mu,\partial_{M_1},0,\dots).
    \label{eq:section_solution_extended}
\end{equation}
Hence, on section one has coordinates of the form $X^{\mathcal M_1}=(y^\mu,X^{M_1},0,\dots)$. This has a direct geometric meaning. The coordinates $y^\mu$ will later be interpreted as fibre coordinates of a principal $H$-bundle, while the $X^{M_1}$ are the extended coordinates familiar from $\mathcal G$-generalised geometry. The present framework therefore combines the principal-bundle structure that we will need for Cartan geometry with the extended tangent structure of generalised geometry.

\section{Derivation from Brane Current Algebra}  \label{chap:CurrentAlgebra}

In the previous section, we introduced an extended generalised geometry that combines a duality group $\mathcal G$ with a local symmetry algebra $\mathfrak h$. We now explain why this structure is natural from the Hamiltonian point of view. More precisely, we show that it arises directly from the current algebra of brane phase space when the latter is written in a $\mathcal G$-covariant form and then supplemented by local $H$-transformations.

The guiding idea is the same as in the point-particle discussion of ordinary Cartan geometry: the relevant geometric structures are encoded in the Poisson algebra of covariant phase-space variables. For branes, however, the phase space is richer. Instead of a single momentum variable, one obtains a hierarchy of currents, and this hierarchy mirrors the tensor hierarchy of generalised geometry.

\subsection{Review: brane currents in \texorpdfstring{$\mathcal{G}$}{G}-generalised geometry}

The relation between $\mathcal G$-generalised geometry and the Hamiltonian description of $\frac12$-BPS $P$-branes has been developed in \cite{Osten:2024mjt}. The basic dictionary is as follows.

A $P$-brane phase space can be organised in terms of currents valued in the representations $\mathcal R_p$ of the tensor hierarchy. Concretely, one obtains a collection of world-volume forms
\begin{equation}
    t_{M_p}\in \mathcal R_p,
\end{equation}
where $t_{M_p}$ is a spatial $(P-p+1)$-form on the brane world-volume. These currents are built from the canonical momentum density, the embedding coordinates, and, where relevant, world-volume gauge fields. Their detailed form depends on the brane under consideration, but the important point is that the tensor hierarchy of $\mathcal G$ is realised directly on phase space.

For example, in the M2-brane case the lowest currents combine the momentum and differentials of target-space coordinates into objects transforming in the appropriate $\mathcal R_p$:
\begin{equation}
        t_{M_1} = (p_m , \mathrm{d}x^m \wedge \mathrm{d}x^{m^\prime}, 0 , \dots) , \quad t_{M_2} = (\mathrm{d}x^m , 0 , \dots) , \quad t_{M_3} = ( 1,0, \dots).
    \end{equation}
The details are not essential here; what matters is that the phase space naturally arranges itself into the same hierarchy that underlies exceptional or doubled geometry.

The Poisson brackets of these currents take a universal form:
\begin{equation}
    \{ t_{A_p}(\sigma),t_{B_q}(\sigma')\}
    =
    -\,\eta^{C_{p+q}}{}_{A_pB_q}\,
    t_{C_{p+q}}(\sigma')\wedge \mathrm d'\delta(\sigma-\sigma').
    \label{eq:CurrentAlgebraSmall_rewrite}
\end{equation}
Up to world-volume boundary terms, this is the brane analogue of the algebraic structure encoded by the $\eta$-symbols of the tensor hierarchy. In particular, the Jacobi identity of this current algebra reproduces the graded Jacobi identities satisfied by the tensor hierarchy product. Thus, the algebra of brane currents is not merely reminiscent of generalised geometry; it realises it directly.

To pass from currents to generalised tensors, one integrates them against coefficient ('test') functions. A section $\Phi$ of an $\mathcal R_p$-bundle is represented by
\begin{equation}
    \Phi=\int \Phi^{A_p}(X(\sigma))\,t_{A_p}(\sigma).
\end{equation}
For such objects, the Poisson bracket reproduces the generalised Lie derivative
    \begin{align}
        \{ \Phi , \Lambda \} = \mathcal{L}_{\Lambda} \Phi,
    \end{align}
for section $\Lambda,\Phi \in \mathcal{R}_1$, provided the currents satisfy, in addition, the brane charge condition
\begin{equation}
    t_{M_p}\wedge \mathrm dX^{N_1}\partial_{N_1}
    =
    D_{M_p}{}^{K_{p-1}N_1}\,t_{K_{p-1}}\partial_{N_1}.
    \label{eq:BraneChargeSmall_rewrite}
\end{equation}
This relation is the bridge between phase-space geometry and target-space generalised geometry. It ensures that the Poisson structure on the brane reproduces the differential structure defined abstractly by the tensor hierarchy.

Finally, the Hamiltonian and diffeomorphism constraints of the brane can be written in terms of the generalised metric and the tensor hierarchy data:
    \begin{align}
        H = \mathcal{H}^{A_1 B_1} t_{A_1} \wedge \star t_{B_1} \approx 0 , \qquad D_{C_2}{}^{A_1 B_1} t_{A_1} \wedge \star t_{B_1} \approx 0.
    \end{align}
In this way, both the kinematics and the dynamics of the brane admit a manifestly $\mathcal G$-covariant description.

The main lesson is therefore that ordinary $\mathcal G$-generalised geometry appears naturally in brane phase space once one organises the canonical variables into the correct hierarchy of currents. We now show how the additional local symmetry $H$ enters the same picture.

\subsection{Adding \texorpdfstring{$\mathfrak h$}{h}-gauge symmetry}\label{sec:addHsym}

Now assume the brane $\sigma$-model has a gauge symmetry $H$. Then the brane current algebra should be supplemented by generators $s_\alpha$ of a local gauge symmetry $\mathfrak{h}$. Their action on the existing brane currents is
\begin{align}
    \{s_\alpha(\sigma),s_\beta(\sigma')\}
    &= f_{\alpha\beta}{}^\gamma s_\gamma(\sigma)\delta(\sigma-\sigma'),
    \label{eq:CurrentAlgebraSeed_rewrite_a}\\
    \{s_\alpha(\sigma),t_{M_p}(\sigma')\}
    &= f_{\alpha M_p}{}^{N_p} t_{N_p}(\sigma)\delta(\sigma-\sigma').
    \label{eq:CurrentAlgebraSeed_rewrite_b}
\end{align}
Thus the $s_\alpha$ generate the local symmetry algebra and act on the tensor hierarchy currents through the representation introduced in the previous section.

At this stage, however, the enlarged algebra is not yet consistent. The Jacobi identity fails unless one introduces additional generators dual to the $H$-currents. Following the mechanism first observed by Pol\'a\v{c}ek and Siegel \cite{Polacek:2013nla} for O$(d,d)$, we add canonical dual one-form currents $\Sigma^\alpha$. The extended algebra then takes the form
\begin{align}
     \{s_\alpha(\sigma),s_\beta(\sigma')\}
     &= f_{\alpha\beta}{}^\gamma s_\gamma(\sigma)\delta(\sigma-\sigma'),
     \nonumber\\
     \{s_\alpha(\sigma),t_{M_p}(\sigma')\}
     &= f_{\alpha M_p}{}^{N_p} t_{N_p}(\sigma)\delta(\sigma-\sigma'),
     \nonumber\\
     \{s_\alpha(\sigma),\Sigma^\beta(\sigma')\}
     &= \delta_\alpha^\beta\,\mathrm d\delta(\sigma-\sigma')
        + f_{\gamma\alpha}{}^\beta \Sigma^\gamma(\sigma)\delta(\sigma-\sigma'),
     \label{eq:CurrentAlgebraFlat_rewrite}\\
     \{\Sigma^\alpha(\sigma),\Sigma^\beta(\sigma')\}
     &=0,
     \qquad
     \{\Sigma^\beta(\sigma),t_{M_p}(\sigma')\}=0,
     \nonumber\\
     \{t_{M_p}(\sigma),t_{N_q}(\sigma')\}
     &=
     -\,\eta^{L_{p+q}}{}_{M_pN_q}\,
     t_{L_{p+q}}(\sigma')\wedge \mathrm d'\delta(\sigma-\sigma')
     \nonumber\\
     &\qquad
     + f_{\alpha M_p}{}^{K_p}\eta^{L_{p+q}}{}_{K_pN_q}\,
     t_{L_{p+q}}(\sigma)\wedge \Sigma^\alpha(\sigma)\delta(\sigma-\sigma').
     \nonumber
\end{align}
The new ingredient is the appearance of $\Sigma^\alpha$ in the current algebra. This is the phase-space signal that a purely $\mathcal G$-covariant description is no longer sufficient once local $H$-transformations are included.

The consistency of this extension is tied to the differential relation
\begin{equation}
    \mathrm dt_{M_p}
    =
    -\,f_{\alpha M_p}{}^{N_p}\,\Sigma^\alpha\wedge t_{N_p},
    \label{eq:Identitydt_rewrite}
\end{equation}
together with the Maurer--Cartan equation for $\Sigma^\alpha$. Conceptually, this means that the currents are no longer closed in the ordinary world-volume sense; rather, their failure to be closed is governed precisely by the local $H$-action. This is the current-algebra counterpart of the twisted fibre dependence that appeared abstractly in the previous section.

Indeed, once the $\Sigma^\alpha$ are identified with the right-invariant Maurer--Cartan forms on $H$,
\begin{equation}
    \Sigma^\alpha = {e_\mu}^\alpha(y)\,\mathrm dy^\mu,
\end{equation}
the origin of the twist becomes clear. The coordinates $y^\mu$ are canonically conjugate to the $H$-generators, and the combination of $t_{M_p}$ with wedge powers of $\Sigma^\alpha$ produces the full hierarchy of extended currents
\begin{equation}
    t^{\alpha_1\dots\alpha_q}_{M_p}
    :=
    t_{M_p}\wedge \Sigma^{\alpha_1}\wedge\dots\wedge\Sigma^{\alpha_q}.
    \label{eq:SigmaComposition_rewrite}
\end{equation}
This is precisely the current-algebra origin of the extended representations introduced previously. From the brane point of view, the objects
\begin{equation}
    t_{\mathcal M_1}
    =
    \big(s_\mu,\;t_{M_1},\;t^\mu_{M_2},\;t^{\mu_1\mu_2}_{M_3},\dots\big)
\end{equation}
span the representation $\mathfrak R_1$, while higher $\mathfrak R_p$ are realised analogously by lower-degree world-volume forms.

In terms of these extended currents, the algebra can be written compactly as
\begin{equation}
     \{ t_{\mathcal M_1}(\sigma),t_{\mathcal N_1}(\sigma') \}
     =
     -\,\eta^{\mathcal L_2}{}_{\mathcal M_1\mathcal N_1}\,
     t_{\mathcal L_2}(\sigma')\wedge \mathrm d'\delta(\sigma-\sigma')
     + f_{\mathcal M_1\mathcal N_1}{}^{\mathcal L_1}\,
     t_{\mathcal L_1}(\sigma)\delta(\sigma-\sigma').
     \label{eq:CurrentAlgebraTwistedModelAlgebra_rewrite}
\end{equation}
This formula is the phase-space realisation of the extended geometry constructed in the previous section. The first term reproduces the extended $\eta$-structure, while the second encodes the local $H$-action and its higher extensions.

As before, sections of the extended bundle are represented by integrating these currents against coefficient functions. Under the corresponding extended brane charge condition, the Poisson bracket reproduces the extended generalised Lie derivative. In this sense, the geometric structure of section~\ref{sec:extGG} is not an abstract invention: it is exactly the structure singled out by brane current algebra once local $H$-symmetry is included.

A useful subtlety should be mentioned. The current algebra itself closes more generally than the minimal extended Lie derivative of the previous section. The restriction of gauge parameters there was needed to obtain strict closure without boundary contributions. On the brane, by contrast, the algebra naturally closes up to total derivatives. This difference is important conceptually: the current algebra sees the full world-volume structure, whereas the geometric formalism isolates the minimal bulk symmetry algebra.

\subsection{The zero-mode algebra} \label{chap:ZeroModeAlgebra}

The current algebra \eqref{eq:CurrentAlgebraTwistedModelAlgebra_rewrite} contains, in its $\delta$-term, structure constants
\[
f_{\mathcal M_1\mathcal N_1}{}^{\mathcal L_1}
\]
that define a Leibniz-type algebra. This algebra will play the role of the model algebra in the later Cartan-geometric construction. We now explain its origin.

The natural objects to consider are the zero-modes of the extended currents. Starting from
\[
t_{A_p}^{\alpha_r}
\equiv
t_{A_p}\wedge \Sigma^{\alpha_1}\wedge\dots\wedge\Sigma^{\alpha_r},
\]
we define
\begin{equation}
    T_{A_p}^{\alpha_r}
    :=
    \int t_{A_p}^{\alpha_r}(\sigma),
\end{equation}
which lies in $\mathfrak R_{p-r}$. These zero-modes inherit two canonical operations from the world-volume structure.

First, there is a differential $Q$ obtained by integrating the world-volume exterior derivative:
\begin{equation}
    QT_{A_p}^{\alpha_r}
    :=
    (-1)^{P+1}\int \mathrm d\,t_{A_p}^{\alpha_r}(\sigma).
    \label{eq:QDef_rewrite}
\end{equation}
Because the world-volume differential squares to zero and because the $H$-action is compatible with the tensor hierarchy, one has
\[
Q^2=0.
\]
Second, there is a graded symmetric product $\bullet$ induced by the $\eta$-symbols,
\begin{equation}
    T_{A_p}^{\alpha_r}\bullet T_{B_q}^{\beta_s}
    =
    (-1)^{1+rq+s}\,
    \eta^{C_{p+q}}{}_{A_pB_q}\,
    T_{C_{p+q}}^{\alpha_r\beta_s}. \label{eq:ZeroModeAlgebra}
\end{equation}
These two operations satisfy the expected compatibility condition:
\begin{equation}
    Q(X\bullet Y)=QX\bullet Y+(-1)^{|X|}X\bullet QY.
    \label{eq:IdentityQCompatibility_rewrite}
\end{equation}
Thus the zero-mode algebra naturally carries the structure of a differential graded algebra, or, after the usual degree shift, a differential graded Lie algebra.

This observation is central. It shows that the extended brane current algebra is not merely a convenient realisation of the previous section, but already contains the higher algebraic structure that will later govern connections and curvatures.

Once the complex is extended by a degree-zero part $\mathfrak R_0$, one may define from $Q$ and $\bullet$ a Leibniz product on $\mathfrak R_1$:
\begin{equation}
    X\circ Y := Y\bullet QX .
\end{equation}
Applied to the zero-modes, this yields precisely the structure constants appearing in the $\delta$-term of the current algebra,
\begin{equation}
    T_{\mathcal M_1}\circ T_{\mathcal N_1}
    =
    f_{\mathcal M_1\mathcal N_1}{}^{\mathcal L_1}T_{\mathcal L_1}.
    \label{eq:genPoincare_rewrite}
\end{equation}
The Leibniz identity follows from the compatibility of $Q$ with $\bullet$ together with the graded Jacobi property of the latter.

This is the algebraic content behind the extended model algebra: it is not postulated, but derived from the zero-modes of brane currents. In this sense, the brane phase space furnishes a concrete and canonical origin for the model algebra that appears later in generalised Cartan geometry.

It is natural to view this structure as a generalisation of the ordinary Poincar\'e or Euclidean algebra. Indeed, in the simplest case $\mathcal G=\mathrm{GL}(d)$ with $H=\mathrm O(d)$, the tensor hierarchy truncates drastically and the above construction reduces to the usual Euclidean algebra. For more general duality groups, one obtains the corresponding higher or generalised analogue. This is why the resulting algebra may be regarded as a generalised Poincar\'e algebra.

To summarise, the extended current algebra of branes provides the physical origin of all the ingredients introduced abstractly in the previous section. The representations $\mathfrak R_p$, the twisted coordinate dependence, the extended generalised Lie derivative, and the Leibniz model algebra all arise naturally from the Hamiltonian structure of brane phase space. This is the reason why the later Cartan-geometric framework is not merely formally consistent, but also physically well motivated.

\section{Curvatures and Torsions in the Linearised Theory} \label{chap:Curvature}

\subsection{Extended Generalised Cartan Geometry}

We now assemble the ingredients developed in the previous sections into a Cartan-geometric framework adapted to extended geometry. The goal is to define curvatures and torsions that are naturally covariant under both generalised diffeomorphisms associated with the duality group $\mathcal G$ and local $H$-transformations. The resulting structure is the natural generalisation of ordinary Cartan geometry to the present setting.

The construction is based on two pieces of data. On the geometric side, one has the extended generalised tangent bundle, defined above as $\mathfrak{R}_1$ , that is naturally associated with the principal $H$-bundle $P\to M$. Locally, it takes the form
\begin{equation}
    \mathfrak R_1 \sim TH \oplus \mathcal R_1[M] \oplus (\mathcal R_2[M]\otimes T^*H)\oplus \dots .
\end{equation}
This is the bundle that combines the local gauge directions with the generalised tangent directions of the physical space. In the $\mathrm O(d,d)$ case it reduces locally to $TP\oplus T^*P$, while in the general case it is organised by the extended hierarchy introduced earlier.

On the algebraic side, one needs a model algebra that plays the role of the homogeneous tangent model. Here, this is not an ordinary Lie algebra, but a differential graded Lie algebra
\begin{equation}
    \mathfrak L = \mathfrak l_0 \oplus \mathfrak{l}_1 \oplus \mathfrak{l}_2 \oplus ...
\end{equation}
equipped with a graded symmetric product $\bullet$ and a differential $Q$. As usual, these data induce on $\mathfrak l_1$ a derived Leibniz product
\begin{equation}
    a\circ b = -Qa\bullet b,
    \qquad a,b\in \mathfrak l_1 .
\end{equation}
The zero-mode algebra \eqref{eq:ZeroModeAlgebra} extracted from brane currents above gives precisely such a structure. We, moreover, assume that $\mathfrak l_1$ contains $\mathfrak h$ as an isotropic Lie subalgebra.

The generalised Cartan connection is then the pointwise identification of these two structures. At degree one it is a bundle map
\begin{equation}
    \theta^{(1)}(p): \ \mathfrak R_1[P]_p\to \mathfrak l_1,
    \label{eq:GeneralisedCartanConnectionDefinition_rewrite}
\end{equation}
and, because it preserves the hierarchy structure, it extends to a family of maps
\begin{equation}
    \theta|(p): \ \oplus_q \mathfrak R_q[P]_p \to \mathfrak L.
\end{equation}
In practice, however, independent information is already contained in $\theta^{(1)}$, and higher maps are fixed by compatibility.

The conceptual role of $\theta$ is the same as in ordinary Cartan geometry: it identifies geometric directions with algebraic generators, while keeping track of the local $H$-symmetry. Accordingly, it satisfies the standard Cartan-type requirements. The vertical directions of the principal bundle are identified with $\mathfrak h$, and the connection obeys the equivariance condition \eqref{eq:Equivariance}. Thus, the local $H$-action is built directly into the definition of the connection.

The generalised Cartan curvature is obtained by evaluating the current algebra on the Cartan connection itself. Writing $\theta$ as a current-valued object, one defines its curvature through
\begin{equation}
       \boxed{
       \{ \theta^{(1)}_{\mathcal A_1}(\sigma),\theta^{(1)}_{\mathcal B_1}(\sigma') \}
       =
       - \eta^{\mathcal C_2}{}_{\mathcal A_1\mathcal B_1}
       \theta^{(2)}_{\mathcal C_2}(\sigma')\wedge \mathrm d' \delta(\sigma-\sigma')
       +
       \Theta_{\mathcal A_1\mathcal B_1}{}^{\mathcal C_1}
       \theta^{(1)}_{\mathcal C_1}(\sigma)\delta(\sigma-\sigma')
       } .
       \label{eq:GeneralisedCartanCurvatureDefinition_rewrite}
\end{equation}
This formula should be understood as the direct analogue of the bracket definition of curvature in ordinary Cartan geometry. The $\delta$-term encodes the structure functions of the twisted model algebra; these are the desired torsions and curvatures. The key difference from the fully enlarged approaches of \cite{Hassler:2023axp} is that here the curvature is defined through the brane current algebra rather than directly through a fully unrestricted generalised Lie derivative. This is precisely what makes the minimal framework possible.

\subsection{Generalised Cartan Connection} \label{chap:Connection}

To describe the connection explicitly, one chooses a local parametrisation adapted to the hierarchy. The appropriate ansatz is a triangular ``extended vielbein'' whose first column contains the independent connection data, while the remaining entries are fixed by compatibility with the hierarchy. Schematically, it takes the form
\begin{equation}
    \theta_{\mathcal A_1}{}^{\mathcal M_1}
    =
    \begin{pmatrix}
    \delta_\alpha^\mu & 0 & 0 & \cdots \\
    \Omega_{A_1}{}^\mu & E_{A_1}{}^{M_1} & 0 & \cdots \\
    \Omega_{A_2}^{\alpha,\mu} & ... & \delta_\mu^\alpha E_{A_2}{}^{M_2} & \cdots \\
    \Omega_{A_3}^{\alpha_1\alpha_2,\mu} & \cdots & \cdots & \ddots
    \end{pmatrix}.
    \label{eq:BigVielbein_rewrite}
\end{equation}
This is the natural generalisation of the ordinary Cartan connection and of the doubled ``megavielbein'' familiar from the $\mathrm O(d,d)$ case \cite{Butter:2021dtu,Hassler:2024hgq}. Its components organise themselves into a tower of connection-like fields. The lowest one, $\Omega_{A_1}{}^\mu$, is a generalised spin connection; higher levels provide additional gauge fields required by covariance.

Not all entries of \eqref{eq:BigVielbein_rewrite} are independent. The connection must preserve the hierarchy structure, in particular the $\eta$-symbols. For examples, linearising this compatibility condition around a fixed generalised frame gives
\begin{equation}
    0
    =
    \delta\theta^{(1)}_{\mathcal A_1}{}^{\mathcal M_1}\eta^{\mathcal C_2}{}_{\mathcal M_1\mathcal B_1}
    +
    \delta\theta^{(1)}_{\mathcal B_1}{}^{\mathcal M_1}\eta^{\mathcal C_2}{}_{\mathcal A_1\mathcal M_1}
    +
    \delta\theta^{(2)}_{\mathcal M_2}{}^{\mathcal C_2}\eta^{\mathcal M_2}{}_{\mathcal A_1\mathcal B_1}.
    \label{eq:variation_rewrite}
\end{equation}
At the linearised level this relation is enough to determine the actual independent fields.

The outcome is simple and important. First, the higher connections $\theta^{(p)}$ for $p>1$ do not introduce new degrees of freedom: they are fixed recursively by $\theta^{(1)}$. Second, among the components of $\theta^{(1)}$, only the first column is independent. Third, those independent components are precisely
\begin{equation}
    \Omega_{M_1}{}^\mu,
    \qquad
    \rho_{M_2}^{\mu_1\mu_2},
    \qquad
    \rho_{M_3}^{\mu_1\mu_2\mu_3},
    \qquad
    \dots
    \label{eq:genConnections_rewrite}
\end{equation}
with the higher fields totally antisymmetric in their $\mathfrak h$-indices. Thus the connection naturally comes as a hierarchy:$\Omega,\ \rho,\ \rho^{(3)},\dots$ rather than as a single one-form connection.

This is one of the central conceptual points of the paper. In ordinary Cartan geometry, curvature covariance requires the coexistence of frame and spin connection. Here the same mechanism repeats at higher levels: the curvature of one connection becomes covariant only after introducing the next connection in the tower. The hierarchy is therefore not an optional embellishment but an intrinsic feature of extended Cartan geometry.

The compatibility conditions also determine the remaining columns of the connection in terms of the basic fields. For example, the first sub-diagonal entries are fixed by
\begin{equation}
    \Omega_{L_{p+1}}^{K_p,\mu}
    =
    -(-1)^p D_{L_{p+1}}{}^{K_pM_1}\Omega_{M_1}{}^\mu.
    \label{eq:ConnectionIdentitySol_rewrite}
\end{equation}
Similarly, higher sub-diagonals are generated from the higher $\rho$-fields. This makes it natural to organise the independent connection data into the parabolic subalgebra $\widetilde{\mathfrak R}_0$, so that, at the linearised level,
\begin{equation}
    \boxed{
    \delta\theta
    =
    \Omega^\mu_{A_1} R^{A_1}_\mu
    +
    \underset{p\geq 2}{\text{\LARGE $\Sigma$}}\frac1{p!}\rho_{A_p}^{\mu_1\dots\mu_p} R^{A_p}_{\mu_1\dots\mu_p}
    } .
    \label{eq:deltathetafromGens_rewrite}
\end{equation}
This expresses the generalised Cartan connection as valued in the same hierarchy algebra that appeared earlier from the zero-mode analysis.

\subsection{A Hierarchy of Curvatures} \label{chap:CartanCurvature}

We now turn to the structure functions $\Theta_{\mathcal A_1\mathcal B_1}{}^{\mathcal C_1}$ that appear in \eqref{eq:GeneralisedCartanCurvatureDefinition_rewrite}. These contain both the twisted model algebra and the physical torsions and curvatures. In the present framework, the most important information is already captured by brackets of the form
\[
\{\theta_\alpha,\theta_{B_p}^{\beta_1\dots\beta_{p-1}}\},
\qquad
\{\theta_{A_1},\theta_{B_p}^{\beta_1\dots\beta_{p-1}}\},
\]
that is, by brackets involving one physical index or one gauge index. The former reproduce the model algebra, while the latter encode the dynamical geometric tensors on the physical space.

When these brackets are decomposed into irreducible pieces, a hierarchy of torsions and curvatures is found. The first family are the $\mathcal R_p$-torsions. At linear order, they depend only on the lowest connection $\Omega_{A_1}{}^\alpha$ and take the universal form
\begin{align}
    \Theta_{A_1B_p}{}^{C_p}{}_{\gamma_1\dots\gamma_{p-1}}^{\beta_1\dots\beta_{p-1}}
    &=
    f_{\alpha B_p}{}^{C_p}\Omega_{A_1}{}^\alpha\,
    \delta^{\beta_1\dots\beta_{p-1}}_{\gamma_1\dots\gamma_{p-1}}
    \nonumber\\
    &\quad
    + f_{\alpha A_1}{}^{D_1}\eta^{L_{p+1}}{}_{D_1B_p}
    D_{L_{p+1}}{}^{C_pE_1}\Omega_{E_1}{}^\alpha\,
    \delta^{\beta_1\dots\beta_{p-1}}_{\gamma_1\dots\gamma_{p-1}}
    \nonumber\\
    &\quad
    - (p-1)f_{[\gamma_{p-1}A_1}{}^{D_1}\delta_{B_p}^{C_p}
    \Omega_{D_1}{}^\alpha
    \delta^{\beta_1\dots\beta_{p-1}}_{\gamma_1\dots\gamma_{p-2}]\alpha}.
    \label{eq:RPTorsion_rewrite}
\end{align}
For $p=1$, this reduces to the familiar generalised torsion of exceptional geometry. In the present setting it should be interpreted as the natural extension of that object to the whole hierarchy.

The second family consists of the curvatures of the connection fields themselves. The lowest one is the curvature of $\Omega$,
\begin{equation}
    \Theta_{A_1B_1}{}^\beta
    =
    -2\partial_{[A_1}\Omega_{B_1]}{}^\beta
    +
    f_{\alpha A_1}{}^{C_1}\eta^{D_2}{}_{C_1B_1}\rho_{D_2}^{\beta\alpha},
    \label{eq:R1Curvature_rewrite}
\end{equation}
while at higher levels one finds, for $p>1$,
\begin{equation}
    \Theta_{A_1B_p}^{\beta_1\dots\beta_p}
    =
    -\partial_{A_1}\rho_{B_p}^{\beta_1\dots\beta_p}
    +
    f_{\alpha A_1}{}^{C_1}\eta^{D_{p+1}}{}_{C_1B_p}
    \rho_{D_{p+1}}^{\beta_1\dots\beta_p\alpha}.
    \label{eq:RPCurvature_rewrite}
\end{equation}
These are the linearised curvatures associated with the hierarchy of connections
\[
\Omega,\rho,\rho^{(3)},\dots .
\]

Their structure makes the geometric mechanism very transparent. The ordinary derivative of a connection is not covariant by itself. Covariance is restored by the next field in the hierarchy. Thus, the curvature of $\Omega$ requires $\rho$; the curvature of $\rho$ requires the next field; and so on. This is exactly the same hierarchical phenomenon that had already appeared in the connection sector, now reflected in the curvature sector.

Many other components of the generalised Cartan curvature are not independent. In particular, the components that carry lower $\mathcal R_q$ labels are largely determined by the above curvatures through the same hierarchy identities that constrain the connection. In this sense, the physically relevant part of the linearised curvature is already captured by the torsions \eqref{eq:RPTorsion_rewrite} and curvatures \eqref{eq:R1Curvature_rewrite}--\eqref{eq:RPCurvature_rewrite}.

It is natural to ask whether these tensors fit into a representation hierarchy. The answer is suggestive, but not completely uniform. The torsions and curvatures may be collected into a space of the form
\begin{equation}
    \mathfrak R_{-1}
    =
    \mathcal R_{-1}
    \underset{q\geq 1}{\text{\Large $\oplus$}}
    \big(\mathcal R_1\otimes \mathcal R_q\otimes \wedge^q\mathfrak h^*\big),
\end{equation}
which plays the role of a curvature representation. For $\mathrm O(d,d)$ this simplifies and aligns particularly neatly with the usual hierarchy, but for general duality groups the corresponding representations lie slightly outside the standard tensor hierarchy.

\paragraph{Ambiguity in the curvature representative.}

A noteworthy feature of the construction is that the curvature is not represented uniquely. Because the current algebra is defined only up to world-volume total derivatives, one may shift the curvature by terms that differ by such boundary contributions. This leads to a one-parameter family of equivalent linearised expressions. A convenient representative is
\begin{align}
    \Theta_{A_1B_1}{}^\beta
    &= -2\partial_{[A_1}\Omega_{B_1]}{}^\beta
    + f_{\alpha A_1}{}^{C_1}\eta^{D_2}{}_{C_1B_1}\rho_{D_2}^{\beta\alpha}
    \nonumber\\
    &\quad
    + \alpha\,\eta^{C_2}{}_{A_1B_1}
    \left(
    f_{\alpha C_2}{}^{D_2}\rho_{D_2}^{\beta\alpha}
    + \frac12 f_{\gamma\delta}{}^\beta \rho_{C_2}^{\gamma\delta}
    \right),
    \label{eq:R1CurvatureGen_rewrite}
\end{align}
and similarly, for $p>1$,
\begin{align}
    \Theta_{A_1B_p}^{\beta_1\dots\beta_{p-1},\beta}
    &=
    -\partial_{A_1}\rho_{B_p}^{\beta_1\dots\beta_{p-1}\beta}
    +
    f_{\alpha A_1}{}^{C_1}\eta^{D_{p+1}}{}_{C_1B_p}
    \rho_{D_{p+1}}^{\beta_1\dots\beta_{p-1}\beta\alpha}
    \nonumber\\
    &\quad
    + \alpha\,\eta^{C_{p+1}}{}_{A_1B_p}
    \left(
    f_{\alpha C_{p+1}}{}^{D_{p+1}}
    \rho_{D_{p+1}}^{\beta_1\dots\beta_{p-1}\beta\alpha}
    + (-1)^{p+1}\frac12
    f_{\gamma\delta}{}^\beta
    \rho_{C_{p+1}}^{\gamma\delta\beta_1\dots\beta_{p-1}}
    \right).
    \label{eq:RPCurvatureGen_rewrite}
\end{align}
Different values of $\alpha$ correspond to different but equivalent representatives, analogous to choosing between more Courant-like and more Dorfman-like presentations. The choice $\alpha=0$ is particularly convenient because it matches the linearised results of \cite{Hassler:2023axp} and fits cleanly into the hierarchy picture.

An important avenue for future work will be to understand the algebraic structure of these curvatures without the detour to current algebras. Potentially, this clear up this ambiguity.

\paragraph{Bianchi identities.} As in ordinary geometry, the curvatures satisfy consistency relations descending from Jacobi identities. The same is to be expected here.

First, Jacobi identities involving one gauge generator $\theta_\alpha = s_\alpha$ imply that all curvature components transform covariantly under $\mathfrak h$. In other words, the curvatures are genuine $H$-tensors. These same identities also relate curvature components in different hierarchy levels.

Second, Jacobi identities involving only physical indices yield differential Bianchi identities. Depending on which representative of the curvature one uses, these may be written in two slightly different ways. In a more ``Courant-like'' presentation one obtains identities in which naked higher connections explicitly appear. This reflects the hierarchical mechanism already emphasised above: the Bianchi identity for one curvature is corrected by the next connection in the tower. For example, the curvature of $\Omega$ is corrected by $\rho$, and its Bianchi identity is in turn corrected by the next higher field. This makes the tensor hierarchy of connections visible not only in the definitions of the curvatures, but also in their consistency conditions.

In a more ``Dorfman-like'' presentation one trades those naked connection terms for additional curvature components, so that the Bianchi identities involve only curvatures. The resulting expressions resemble Leibniz identities more closely and are often algebraically cleaner, though less economical in the number of curvature objects that must be introduced.

From the present perspective, both interpretations are useful. The first highlights the iterative role of the higher connections in restoring covariance. The second emphasises the algebraic closure properties of the curvature hierarchy. In either formulation, the main message is the same: extended generalised Cartan geometry naturally organises torsions, curvatures, and their Bianchi identities into a hierarchy governed by the same higher-algebraic structure that already appeared in the current algebra and in the connection sector.

\section{Outlook}\label{chap:Outlook}

In this article, we developed a \emph{minimal} framework for gauged exceptional geometry, namely an extended geometry that makes manifest both generalised diffeomorphisms together with their tensor hierarchy and an additional compatible gauge symmetry. In this way, we revisited and extended earlier ideas on generalisations of Cartan geometry \cite{Polacek:2013nla,Hassler:2022egz,Hassler:2023axp,Hassler:2025qhh} in \cite{Hassler:2024hgq,Hassler:2025rag}, and proposed a setting in which curvatures can be defined covariantly with respect to both structures. As a first test of the construction, we derived in section~\ref{chap:CartanCurvature} the linearised expressions for the basic torsions and curvatures in the hierarchy. In particular, the $\mathcal{R}_p$-curvatures and the $\mathcal{R}_p$-torsion emerge as the independent contributions to the generalised Cartan curvature. Even at the linearised level, the resulting structure already displays qualitatively new features, such as the hierarchy of higher connections and the appearance of both Dorfman- and Courant-like versions of curvatures and Bianchi identities. The obvious next step is therefore to extend these results to the full non-linear theory.

Pursuing this problem further will require a more complete control over the tensor hierarchy itself, in particular explicit expressions for the relevant $\eta$- and $D$-symbols across all representations $\mathcal{R}_p$. At the same time, the formalism developed here strongly suggests that the index-heavy presentation should admit a more intrinsic reformulation. Since our construction is naturally organised by a differential graded Lie algebra, one expects an index-free description closer in spirit to the standard differential-form language of Cartan geometry. This also points to a natural generalisation: differential graded Lie algebras are only the strict subclass of $L_\infty$ algebras, so one is led to ask whether the present framework can be extended to genuine higher brackets \cite{Lada:1992wc,Hohm:2017pnh,Borsten:2021ljb}. Such a generalisation is not merely formal. In realistic tensor hierarchies for $E_{d(d)}$, the simple graded symmetry pattern breaks down at sufficiently high degree, and this is one of the main obstacles for the \emph{maximal} approach of \cite{Hassler:2023axp}, where $H\times\mathcal G$ is embedded into a larger exceptional symmetry. In practice, that strategy imposes strong restrictions on the allowed dimension of the gauge group $H$. By contrast, the minimal construction proposed here avoids this obstruction, although it does so at the price of no longer incorporating the enlarged generalised U-dualities studied in \cite{Hassler:2025qhh}. On the other hand, precisely because gauge groups $H$ of arbitrary dimension can now be treated, the present framework opens the door to several geometric questions that were previously difficult to formulate.

A first issue concerns metric compatibility. In the present work, we concentrated on the construction of covariant torsions and curvatures. For physical applications, however, one also needs a suitable notion of compatibility between the connection and the generalised metric, since this is essential for identifying the propagating degrees of freedom. It is already well known in $\mathrm O(d,d)$ generalised geometry that the analogues of torsion-freeness and metric compatibility do not uniquely determine the generalised affine connection \cite{Hohm:2012mf}. The same phenomenon persists in the Cartan-geometric setting considered here. One possible way forward is to introduce an additional gauge symmetry that removes the unfixed components, as proposed in \cite{Butter:2021dtu}. Yet this symmetry does not close by itself and requires a further extension. In the $\mathrm O(d,d)$ case this leads to a graded, infinite-dimensional gauge group $H$, and carrying out the analogous construction in M-theory seems to require precisely the sort of formalism developed in the present paper. Closely related infinite-dimensional gauge groups also appear in higher-derivative corrections through the generalised Bergshoeff--de Roo mechanism \cite{Baron:2018lve,Baron:2020xel}, as was recently emphasised in \cite{Gitsis:2024gfb}.

A second direction concerns the interpretation of the resulting hierarchy of connections and curvatures. The structures found here bear a strong resemblance to those of higher gauge theory \cite{Baez:2010ya,Grutzmann:2014hkn,Ritter:2015zur,Borsten:2021ljb,Borsten:2024gox}. Recasting the present construction in that language therefore appears both feasible and conceptually promising. In particular, one may hope that extended generalised Cartan geometry points toward a gravitational counterpart of higher gauge theory, much as ordinary Cartan geometry provides a gauge-theoretic perspective on gravity. From that point of view, the hierarchy of higher connections uncovered here may not be an auxiliary technical device, but part of the correct geometric structure underlying duality-covariant gravity theories.

Finally, brane current algebras \cite{Alekseev:2004np,Bonelli:2005ti,Hatsuda:2012uk,Hatsuda:2012vm,Hatsuda:2013dya,Hatsuda:2020buq,Osten:2021fil,Arvanitakis:2021wkt,Hatsuda:2023dwx,Osten:2024mjt} played an essential role throughout this work, mainly as a tool for deriving the relevant geometric objects. But their role is likely to be more than merely technical. As already suggested in \cite{Osten:2024mjt}, the gauged exceptional geometry introduced here should be particularly relevant for exotic branes, such as the Kaluza--Klein monopole, where transverse directions may carry non-trivial gauge symmetries. Early indications in this direction were given in \cite{Sakatani:2017vbd}. It is plausible that the world-volume theories of such branes can be constructed systematically from gauged exceptional geometry together with the current-algebraic methods of \cite{Osten:2024mjt}. Such a development would provide a more systematic framework for analysing their dynamics and would at the same time offer a natural setting in which U-dualities and their generalisations \cite{Sakatani:2019zrs,Malek:2019xrf} can be studied directly on the world volume.

\section*{Acknowledgements}
The speaker thanks Martin Cederwall, Axel Kleinschmidt and Josh O'Connor for discussions, and Falk Hassler, Ondra Hulik, Yuho Sakatani and Alex Swash for collaboration on these topics. The research of D.~O.~was supported by SONATA grant No. 2024/55/D/ST2/01205 funded by NCN.

\end{document}